\documentclass[10pt,twocolumn,twoside]{IEEEtran}

\usepackage{cite}
\usepackage{latexsym}
\usepackage{verbatim}

\usepackage{amsfonts}
\usepackage{epsfig}
\usepackage[cmex10]{amsmath}
\usepackage{color}
\usepackage{epsfig}
\usepackage{url}
\usepackage{cite}
\usepackage{enumerate}
\usepackage{epstopdf}
\usepackage{amssymb}
\usepackage{graphicx}
\usepackage{amsmath}
\usepackage{graphicx}
\usepackage{marvosym}
\usepackage{steinmetz}
\usepackage{amsthm}
\usepackage{xcolor}
\usepackage{cancel}
\usepackage{soul}
\usepackage{soul}
\usepackage{changes}
\usepackage{algorithm}
\usepackage{algpseudocode}
\usepackage{lipsum}
\definechangesauthor[name={Per cusse}, color=orange]{per}
\setremarkmarkup{(#2)}

\renewcommand{\baselinestretch}{0.96}
\newtheorem{lem}{Lemma}

\allowdisplaybreaks
\newcommand{\norm}[1]{\left\lVert#1\right\rVert}

\newcommand*{\suchthat}{\;\ifnum\currentgrouptype=16 \middle\fi|\;}
\allowdisplaybreaks

\begin{document}

\title{Secure State Estimation and Control for Cyber Security of AC Microgirds}

\author{Dariush Fooladivanda*, Qie Hu*, Young Hwan Chang*, and Peter W. Sauer
\thanks{D. Fooladivanda is with the Department of Mechanical and Aerospace Engineering at the University of California, San Diego, USA (dfooladi@ucsd.edu). Q. Hu is with the Department of Electrical Engineering and Computer Sciences, University of California Berkeley, USA (qiehu@berkeley.edu). Y. H. Chang is with the Department of Biomedical Engineering and Computational Biology Program, Oregon Health and Science University, USA (chanyo@ohsu.edu). Peter W. Sauer is with the Department of Electrical and Computer Engineering, University of Illinois Urbana-Champaign, USA (psauer@illinois.edu).}
\thanks{*These authors contributed equally to this work.}}

\markboth{}
{Shell \MakeLowercase{\textit{et al.}}: Bare Demo of IEEEtran.cls for Journals}

\maketitle
\begin{abstract}
A timely, accurate, and secure dynamic state estimation is needed for reliable monitoring and efficient control of microgrids. The synchrophasor technology enables us to obtain synchronized measurements in real-time and to develop dynamic state estimators for real-time monitoring and control of microgrids. 
In this study, we consider an AC microgrid comprising several synchronous generators and inverter-interface power supplies, and focus on securely estimating the dynamic states of the microgrid from a set of corrupted data. We propose a dynamic state estimator which enables the microgrid operator to reconstruct the dynamic states of the microgrid from a set of corrupted data. Finally, we consider an AC microgrid with the same topology as the IEEE 33-bus distribution system, and numerically show that the proposed secure estimation algorithm can accurately reconstruct the attack signals.
\end{abstract}
\begin{IEEEkeywords}
\textcolor{black}{Secure state estimation, dynamic state estimation, microgrids, distribution systems.}
\end{IEEEkeywords}
\IEEEpeerreviewmaketitle



\section{Introduction}
\IEEEPARstart{F}{uture} distribution systems will include interconnected microgrids \cite{ref1}-\!\!\cite{ref2}. Such systems are designed to be operated and controlled in a hierarchical fashion to deal with various dynamic phenomena with different time scales \cite{ref_new_1}-\!\!\cite{ref_new_2}. The functionality of control systems in microgrids is highly dependent on state estimation schemes used to observe the state of the system over time. Traditional energy management systems are using static state estimation schemes designed from steady-state models \cite{ref_new_3}-\!\!\cite{ref_new_4}. With the high penetration of distributed energy resources (DER) on the generation side, and flexible loads and new demand-response technologies on the demand side, static state estimation methods will be unable to capture power systems' dynamics accurately \cite{ref_new_12}. Dynamic state estimation schemes will be needed to accurately capture the system dynamics. Such state estimation schemes, designed based on dynamical models, will play an important role in microgrid control and protection \cite{ref_new_5}-\!\!\cite{ref_new_6}. The focus of this paper is the design of dynamic state estimation schemes which are secure to cyber-physical attacks.

Large-scale integration of inverter-interfaced power supplies and distributed controls requires a widespread deployment of the synchrophasor technology and communication networks in future distribution systems. This will lead to high-volume data exchange between different controllers which will make microgrids vulnerable to cyber-physical attacks. Corrupted data in controllers can disrupt power generators' synchrony and result in a network-wide instability. Several attack detection schemes have been proposed in recent years \cite{ref_new8}-\!\!\cite{ref_new13}. In false-data injection attacks, disturbances are injected to sensors and actuators to disrupt their measurements and computed control inputs. While it is possible to identify misbehaving agents \cite{ref_new8}-\!\!\cite{ref_new9}, such solutions require knowledge of the communication network and are hard to scale. In \cite{ref_new14}, the authors propose a computationally-efficient scheme to detect deception attacks on sensors. While such solutions enable us to detect cyber attacks efficiently, these solutions do not mitigate all possible adverse effects. Robust game-theoretic schemes can lead to conservative results against cyber attacks \cite{ref_new15}-\!\!\cite{ref_new16}.

Cyber attacks can be modeled as noise or disturbance to the system. Basseville \textit{et al.} \cite{ref_new19} use noise filtration techniques to detect and remove malicious attacks, and the authors in \cite{ref_new20} develop disturbance attenuation methods for cyber attack detection. Notice that noise filtration or disturbance attenuation techniques operate under certain statistical properties, e.g., white Gaussian noise signal. However, in practice, cyber attacks can be deliberately designed by a malicious attacker. In \cite{ref_new_7}, the authors propose an attack-resilient distributed control for synchronization of islanded inverter-based microgrids. The authors study the effect of attacks on sensors and actuators, and numerically demonstrate the effectiveness of their distributed controls on a modified IEEE 34-bus system. 

\textcolor{black}{In recent years, several researchers have focused on state estimation with security guarantees for both linear and nonlinear dynamical systems \cite{ctrl_sec8}-\!\!\cite{ctrl_nonlin}. The existing literature related to secure state estimation in linear dynamical systems can be classified into two classes: 1) noiseless measurements, and 2) noisy measurements. For the noiseless framework, the studies in \cite{ctrl_sec8}-\!\!\cite{ctrl_sec9} show that the state of the system can be accurately reconstructed under certain conditions. When sensor measurements are affected by both noise and attack signals, sparse attack vectors can be distinguished from measurement noise under certain conditions \cite{new5}-\!\!\cite{new10}. In \cite{ctrl_nonlin}, the authors focus on securely estimating the state of a nonlinear dynamical system from a set of corrupted measurements for two classes of nonlinear systems without any assumption on the sensor attacks or corruptions. They propose a secure state estimator for those systems assuming that the set of attacked sensors can change with time. Finally, they consider a power system comprising several synchronous generators and storage units connected to each other through a set of transmission lines, and assume that the storage units use feedback linearization controls. The authors design a secure estimator that allows us to securely estimate the dynamic states of the power network, i.e., the states of the synchronous generators, under cyber-physical attacks and communication failures.}

Several approaches in the literature are addressing the dynamic state estimation problem. However, the existing dynamic state estimators have the following drawbacks:
\begin{enumerate}
\item Loads are considered to be quasi-static, i.e., load dynamics are neglected. Unlike large power systems, microgrids are small footprint power systems comprising distribution assets, DERs, and loads. DERs are typically of inverter-interfaced power supplies with no inertia. These resources respond to disturbances as fast as their controls, and hence dynamics of loads and DERs are strongly coupled in microgrids. More precisely, load dynamics have significant transient impacts in microgrids \cite{load_dyn1}-\!\cite{load_dyn2}.
\item Controllers are considered to have specific structure. For example, storage units are assumed to use feedback linearization controls in \cite{ctrl_nonlin}. DERs can use several types of distributed and centralized  control schemes \cite{ctrl_dyn2}-\!\!\cite{ctrl_dyn3}. 
\end{enumerate}
In summary, load dynamics cannot be neglected in securely estimating dynamic states, and controllers cannot be limited to any specific structure. To overcome the above drawbacks, we develop a secure state estimation method without linearization or reducing the microgrid into a network of oscillators.



In this study, we focus on secure dynamic state estimation in AC microgrids under cyber attacks or communication failures. We consider an AC microgrid comprising several synchronous generators, inverter-interfaced power supplies, and loads controlled via a central controller (i.e., the microgrid operator) as well as local controllers. We assume that these controllers use the synchrophasor technology, phasor measurement units (PMU), to maintain the system's reliability. More precisely, each bus is equipped with a local controller, transceiver, and measurement unit which allow the controller to exchange its information with the central controller and the local controllers of other buses. The transceivers are connected through a communication network which sends the feedback information to the microgrid operator. We make the following assumptions:
\begin{itemize}
\item[A.1] The communication paths from the central controller to different buses are secured while other communication paths and PMUs are subject to cyber attacks. 
\item[A.2] The set of attacked PMUs or communication paths can change with time. 
\item[A.3] Attacks or corruptions can follow any particular model. 
\end{itemize}
A practical example of cyber attacks in which the set of attacked sensors can change with time is provided in \cite{kundur_time}.

We propose a secure state estimator for reconstructing the dynamic states of the AC microgrid from a set of corrupted measurements without any assumption on the attacks. The proposed estimators are computationally efficient. We then consider an AC microgrid with the same topology as the IEEE 33-bus distribution system, and numerically demonstrate the effectiveness of our estimators in accurately reconstructing the attack or corruption signals. The proposed technique enables microgrid operators to reconstruct dynamic states before using the received measurements for computing control signals, and to monitor the operation of local controllers.

The paper is organized as follows: In Section \ref{sys_sec}, we introduce the microgrid model adopted in this work, and review the classical error correction methods. In Section \ref{sec_nonlinear_estimation}, we formulate the dynamic state estimation problem for AC microgrids, and propose a solution technique for recovering dynamic states. Finally, in Section \ref{num_sec}, we demonstrate the effectiveness of the proposed state estimation algorithm using the IEEE 33-bus distribution
system.

\section{Preliminaries}\label{sys_sec}
We first introduce the physical layer model of an AC microgrid with synchronous generators, inverter-interfaced power supplies, and loads. We then describe the cyber layer over which the microgrid operator relies to control its
DERs, and introduce the cyber attacks that we are considering in this study. Finally, we briefly review the existing secure state estimation techniques for linear dynamical systems.


\subsection{Physical Layer Model}
Consider a microgrid with $m+n+l$ buses. Without loss of generality, we assume that $m$ of these buses, indexed by $\mathcal{N}^{(S)}=\{1,\cdots,m\}$, have synchronous generators connected to them, and that $n$ of the microgrid buses, indexed by $\mathcal{N}^{(I)}=\{m+1,\cdots,m+n\}$, have inverter-interfaced power supplies attached to them. We further assume that the remaining buses, indexed by $\mathcal{N}^{(L)}=\{m+n+1,\cdots,m+n+l\}$, have loads and no generation\footnote{Without loss of generality, we consider tie-line buses that connect the microgrid to the grid, as load buses.}. In addition, we assume that the network interconnecting the generator and load buses is linear so that it can be represented by the nodal admittance matrix $Y=G+\sqrt{-1} B$ where $G$ and $B$ denote the conductance and susceptance matrices, respectively.

We use the standard structure-preserving model to describe the microgrid's dynamics \cite{structure_prev}. This model that incorporates the dynamics of generators' rotor angle and response of load power output to frequency deviations, allows us to preserve the structure of the grid, i.e., no load bus elimination is made. We introduce fictitious buses representing the internal generation voltages for synchronous generators and inverter-interfaced power supplies. Each of these buses is connected to either a synchronous generator or inverter-interfaced power supply bus via reactances that account for transient reactances and connecting lines. These reactances can be considered as transmission lines \cite{sauer}. Therefore, in the augmented network, the number of buses is $2(m+n)+l$. We further denote the set of fictitious buses for synchronous generators and inverter-interfaced power supplies by $\mathcal{N}_f^{(S)}$ and $\mathcal{N}_f^{(I)}$, respectively. The topology of the augmented network can be represented by an undirected graph, $\mathcal{G}(\mathcal{N},\mathcal{E})$, where $\mathcal{N}=\mathcal{N}^{(S)}\cup\mathcal{N}^{(I)}\cup\mathcal{N}^{(L)}\cup\mathcal{N}_f^{(S)}\cup\mathcal{N}_f^{(I)}$; and where each element in the edge set $\mathcal{E}$ corresponds to a transmission line connecting a pair of buses in the augmented network. We assume that the network topology is fixed and known to the microgrid operator.

Let $V_i$ and $\theta_i$ denote the voltage magnitude and phase angle of bus $i\in\mathcal{N}$, respectively. The dynamics of each synchronous generator are described by a structure-preserving model with constant complex voltage
behind reactance \cite[Sec. 7.9.2]{sauer}, augmented to include the governor dynamics. For a synchronous generator at fictitious bus $i\in\mathcal{N}_f^{(S)}$, let $\theta_i$ denote the angle of its voltage (the generator's internal voltage) as measured with respect
to a synchronous reference rotating at the nominal system electrical frequency $\omega_0$. Further, let $\omega_i$ denote its rotor electrical
angular speed, and let $P_i^m$ be the turbine's mechanical power. For each synchronous generator $i\in\mathcal{N}_f^{(S)}$, we have
\begin{align}
\label{swing1k}\dot{\theta}_i&=\omega_i-\omega_0,\\
{M_i}\dot{\omega}_i&=P_i^m-{D_i}(\omega_i-\omega_0)\nonumber\\
\label{swing2k}&\quad\quad-\sum_{j\in\mathcal{N}_i}{{V_i V_j |y_{ij}|~\text{sin}(\theta_i-\theta_j+\phi_{ij})}},\\
\tau_i \dot{P_i^m}&=-P_i^m+P_i^s-R_i (\omega_{i,{\text{meas}}}-\omega_0),
\end{align}
where $y_{ij}=g_{ij}+\sqrt{-1} b_{ij}$ with $g_{ij}$ and $b_{ij}$ being the entries of the conductance and susceptance matrices, respectively, and $\phi_{ij}$ equals $\text{arctan}(g_{ij}/b_{ij})$. $\mathcal{N}_i$ denotes the set of neighbors of node $i$ in graph $\mathcal{G}(\mathcal{N},\mathcal{E})$, $D_i$ (in s/rad) is the generator damping coefficient, and $M_i$ (in $s^2$/rad) is a scaled inertia constant. Further, $R_i$ is the frequency-power speed-droop characteristic constant, $\tau_i$ is the generator governor time constant, and $P_i^s$ denotes the generator's power set-point. Notice that $\omega_{i,{\text{meas}}}$ is the measured value of state $\omega_i$ for all $i\in\mathcal{N}_f^{(S)}$.

The dynamics of inverter-based power supplies can similarly be described by a constant voltage behind reactance model, augmented to include a frequency-droop controller. For a inverter-based power supply at fictious bus $i\in\mathcal{N}_f^{(I)}$, let $\theta_i$ denote the angle of its voltage as measured with respect
to the nominal system electrical frequency $\omega_0$, and let $P_i^s$ denote its generation set-point. For each inverter-based power supply $i\in\mathcal{N}_f^{(I)}$, we have
\begin{align}
\label{swing1kI}D_i\dot{\theta}_i&=P_i^s-\sum_{j\in\mathcal{N}_i}{{V_i V_j |y_{ij}|~\text{sin}(\theta_i-\theta_j+\phi_{ij})}},
\end{align}
where $D_i$ (in s/rad) is the speed-droop characteristics slope of the inverter-interfaced power supply.

In general, the real power drawn by the load at load bus $i\in\mathcal{N}^{(S)}\cup\mathcal{N}^{(I)}\cup\mathcal{N}^{(L)}$ is a nonlinear function of voltage and frequency. However, for constant voltages and small frequency deviations around the nominal frequency $\omega_0$, it is
reasonable to assume that the real power drawn by the load equals ${P}_i^d+D_i \dot{\theta_i}$ where $D_i > 0$ and ${P}_i^d>0$ are the constant frequency coefficient of load and the nominal load at bus $i\in\mathcal{N}^{(S)}\cup\mathcal{N}^{(I)}\cup\mathcal{N}^{(L)}$, respectively. Therefore, for constant voltages and small frequency deviations, the dynamics of the phase angle at load bus $i\in\mathcal{N}^{(S)}\cup\mathcal{N}^{(I)}\cup\mathcal{N}^{(L)}$ can be modeled by
\begin{align}
\label{load_bus}&D_i \dot{\theta_i}=-{P}_i^d-\sum_{j\in\mathcal{N}_i}{{V_i V_j |y_{ij}|~\text{sin}(\theta_i-\theta_j+\phi_{ij})}}.
\end{align}
Here, we ignore reactive powers for synchronous generators, inverter-interfaced power supplies, and loads as they are unnecessary for the analysis presented in this study.

\subsection{Cyber Layer Model}
Each bus $i$ is equipped with a local controller, transceiver, and measurement unit which allow bus $i$ to exchange information with the central controller (i.e., the microgrid operator) as well as the local controllers of other buses. As mentioned earlier, fictitious buses represent the internal generation
voltages for synchronous generators and inverter-interfaced power supplies. We assume that each inverter-interfaced power supply is equipped with a local controller, transceiver, and measurement unit which allow the inverter-interfaced power supply at fictitious bus $i$ to measure and communicate its internal voltage phase angle.

The rotor angle of synchronous generators has an electro-mechanic nature, and hence it cannot be measured via electric measurement units. However, we can estimate rotor angle by using other electric parameters of synchronous generators. In the literature, several studies are addressing the problem of estimating the rotor angle of synchronous generators \cite{rotor1}-\!\cite{rotor4}. We assume that each synchronous generator is equipped with a local controller, transceiver, and measurement unit, and that the measurement unit at bus $i$ is using a rotor angle estimator to compute the internal voltage phase angle for the synchronous generator at bus $i$.

\begin{figure}
\includegraphics[width=0.42\textwidth]{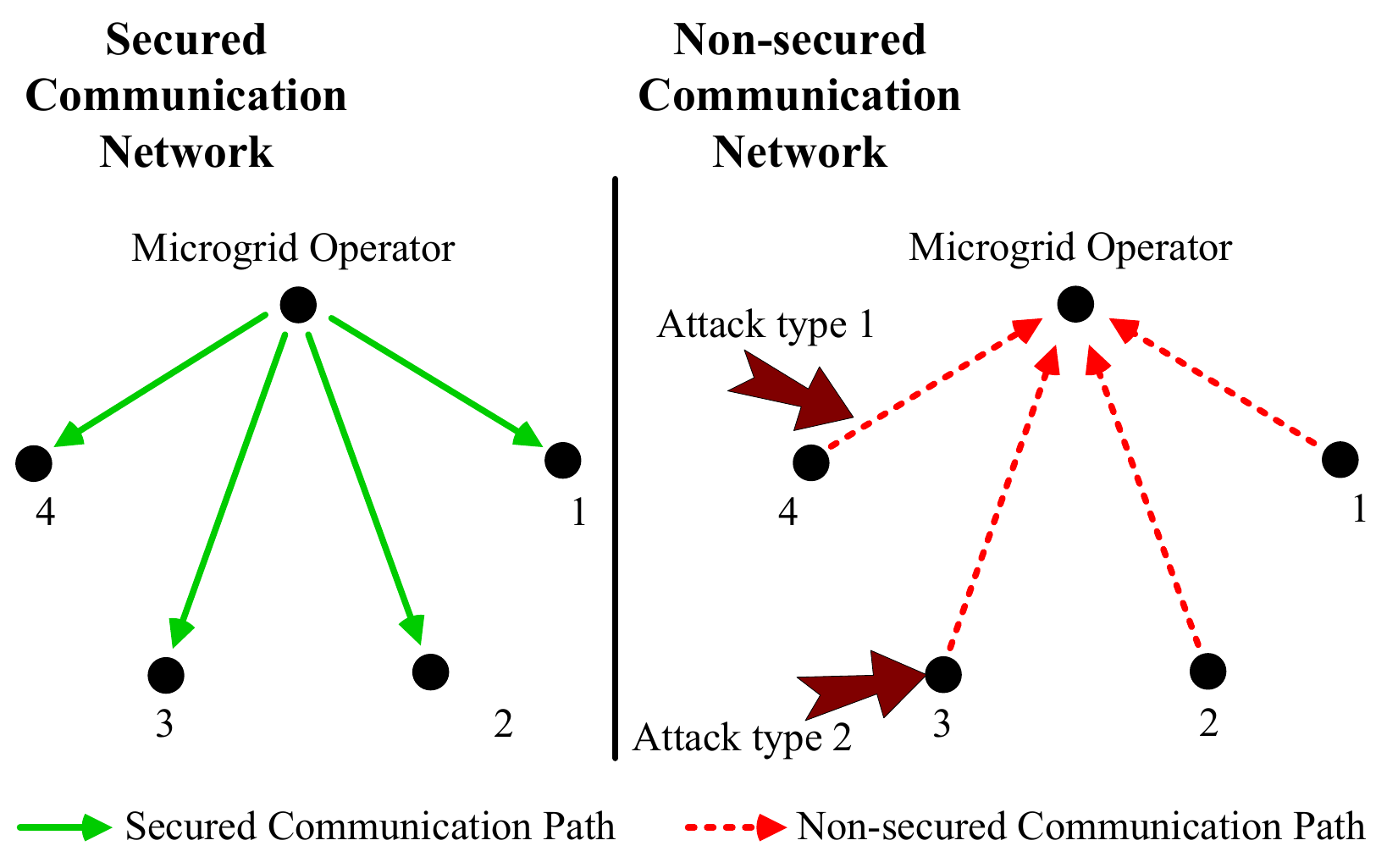} \caption{A graphical depiction of the cyber network model: For simplicity, we focus on a microgrid comprising four buses. The graph in green (sold lines) shows the secured information flow (i.e., the set of information flows from the microgrid operator to local controllers) while the graph in red (dotted lines) represents the set of non-secured information flows.}
\label{wacs}
\end{figure}

The transceivers are connected through a communication network which sends the feedback information measurements, including rotors' speeds and voltage phase angles, to the central controller. The communication network, measurement units, and transceivers are not secured. Therefore, these devices are subject to cyber attacks and communication failures. We assume that the communication paths from the central controller to the local controllers are secured while other communication paths are not secured\footnote{In practice, the central controller is strongly protected against cyber attacks, and the local transceivers are more vulnerable to cyber attacks than the central controller. For more information, we refer the reader to the North American Electric Reliability Corp. (NERC)s Critical Infrastructure Protection (CIP) standards \cite{nerc}.}. Hence, the communication network can be decomposed into two sub-graphs, one for secured information flow and one for non-secured information flow, as shown in Fig.~\ref{wacs}.

\textcolor{black}{The microgrid operator needs to perform secure state estimation before using the received measurements (e.g., $\omega_i$'s and $\theta_i$'s) for computing different control signals and for monitoring the performance of the local controllers in the microgrid. To do so, we consider two types of attacks:
\begin{itemize}
\item \textbf{Attack 1:} an attack that corrupts the communication paths from the transceivers to the central controller.
\item \textbf{Attack 2:} an attack that affects measurement units.
\end{itemize}
These types of attacks are illustrated in Fig.~\ref{wacs} for a microgrid comprising four buses. Notice that the set of attacked measurements can change at each time instant.}



\subsection{Error Correction}\label{sec_review}
\subsubsection{Compressed Sensing} Let $A \in \mathbb{R}^{m\times n}~ (m \ll n)$ and $b \in \mathbb{R}^m$ be a sensing matrix and measurement vector. Sparse solutions $x\in \mathbb{R}^n$, are sought to the following problem:
\begin{equation}
	\min_x \norm{x}_0~\text{subject to}~b= Ax,
	\label{eq:CS}
\end{equation}
where $\norm{x}_0$ denotes the number of nonzero elements of $x$. The following results provide a sufficient condition for a unique solution to (\ref{eq:CS}).

\begin{lem} (\hspace{1sp}\cite{Candes_Tao}) \label{lem:CS}
If the sparsest solution to (\ref{eq:CS}) has $\norm{x}_0 = q$ and $m\ge 2q$ and all subsets of $2q$ columns of $A$ are full rank, then the solution is unique.
\end{lem}

\subsubsection{The Error Correction Problem} 

Consider a full rank coding matrix $C\in \mathbb{R}^{\ell\times n}$ $(\ell > n)$ and a set of corrupted measurements $y=Cx + e$ where $e$ is an arbitrary and unknown sparse error vector. In the classical error correction problem, we aim to recover the input vector $x \in \mathbb{R}^n$ from the corrupted measurements $y$. To reconstruct $x$, it is sufficient to reconstruct the error vector $e$ since knowledge of $y=Cx + e$ and $e$ gives $Cx$, and then $x$ can be computed using the full rank matrix $C$ \cite{Candes_Tao}. Candes \textit{et al.} \cite{Candes_Tao} construct a matrix $F$ that annihilates $C$ on the left, i.e.,  $FCx = 0$ for all $x$. Then, by applying $F$ to $y$, they obtain
\begin{equation}
	\tilde y = F (Cx + e) = Fe.
\end{equation}
Hence, the classical error correction problem can be transformed into reconstructing a sparse vector $e$ from the measurement vector $\tilde y = Fe$. By using Lemma \ref{lem:CS}, if all subsets of $2q$ columns of $F$ are full rank, we can then reconstruct any $e$ that satisfies $\| e \|_0 \leq q$.


\subsubsection{Secure Estimation for Linear Dynamical Systems} Consider a discrete-time linear control system as follows:
\begin{equation}
\begin{aligned}
&x[k+1]= A x[k] ,\\
&y[k] = C x[k] + e[k],
\end{aligned}
 \label{eq:system_model_se}
\end{equation}
where $x[k] \in \mathbb{R}^n $ and  $y[k] \in \mathbb{R}^p$ denote the states and outputs of the system at time slot $k$, respectively. $e[k] \in \mathbb{R}^p$ denotes attack signals injected by malicious agents at the sensors.

Let $Y \in \mathbb{R}^{p\cdot K}$ be a collection of corrupted measurements over $K$ time slots, and let $E_{q,K}$ denote the set of error vectors $\begin{bmatrix} e[0]; ~ ...~  ;  e[K-1] \end{bmatrix}   \in  \mathbb{R}^{p\cdot K} $ where each $e[k]$ satisfies $\|e[k]\|_0 \leq q \leq p$. We have:
\begin{eqnarray} \label{eq:sys_err_corr}
\begin{aligned}
	Y &\triangleq \begin{bmatrix} y[0] \\ y[1] \\ \vdots \\ y[K-1] \end{bmatrix}
		= \begin{bmatrix} Cx[0] + e[0]\\ CA x[0] + e[1] \\ \vdots \\ CA^{K-1} x[0] + e[K-1] \end{bmatrix} \\
		& =
		\begin{bmatrix} C \\ CA \\ \vdots \\ CA^{K-1} \end{bmatrix} x[0] + E_{q,K} \triangleq \Phi x[0] + E_{q,K},
		\label{eq:decoder_Phi}
\end{aligned}
\end{eqnarray}
where  $\Phi \in \mathbb{R}^{p\cdot K \times n}$ is the $K$-step observability matrix of the system at hand. We assume that $\operatorname{rank}(\Phi) = n$; otherwise, we cannot determine $x[0]$ even if there was no attack $E_{q,K} = 0$.

We can now reconstruct the initial state $x[0]$ from $y[k]$'s, where $k=0,...,K-1$, by following a two-step procedure \cite{Candes_Tao} and \cite{David_Chang2}. First, we compute the error vector $E_{q,K}$, and then solve for $x[0]$. To compute $E_{q,K}$, consider the $QR$ decomposition of $\Phi$ as follows:
\begin{eqnarray}
	\Phi = \begin{bmatrix} Q_1 & Q_2 \end{bmatrix} \begin{bmatrix} R_1 \\ 0 \end{bmatrix} = Q_1 R_1,
\end{eqnarray}
where $\begin{bmatrix} Q_1 & Q_2 \end{bmatrix} \in \mathbb{R}^{p\cdot K \times p\cdot K}$ is orthogonal, $Q_1 \in \mathbb{R}^{p\cdot K\times n}, Q_2 \in \mathbb{R}^{p\cdot K \times (p\cdot K-n)}$, and $R_1 \in \mathbb{R}^{n\times n}$ is a rank-$n$ upper triangular matrix.
We pre-multiply (\ref{eq:decoder_Phi}) by $\begin{bmatrix} Q_1 & Q_2 \end{bmatrix} ^\top$, and obtain:
\begin{equation}
	\begin{bmatrix} Q_1 ^\top \\ Q_2 ^\top \end{bmatrix} Y = \begin{bmatrix}R_1 \\ 0  \end{bmatrix} x[0]+ \begin{bmatrix} Q_1 ^\top \\ Q_2^\top \end{bmatrix} E_{q,K}.
	\label{eq:QR}
\end{equation}
Using the second block row, we have:
\begin{equation}
	\tilde Y \triangleq Q_2^\top Y = Q_2^\top E_{q,K},
	\label{eq:E_est}
\end{equation}
where $Q_2^\top \in \mathbb {R} ^{ (p\cdot K-n) \times p\cdot K}$. \textcolor{black}{Using Lemma \ref{lem:CS}, the above optimization has a unique, $s$-sparse solution (where $s\le q\cdot K$) if all subsets of $2s$ columns (at most $2 q\cdot K$ columns) of $Q_2^\top$ are full rank. Using this observation, we consider solving the following $l_1$-minimization problem:}
\begin{equation}
	\hat{E}_{q,K} = \arg \min_E \norm { E}_{l_1} \text{ subject to } \tilde Y = Q_2^\top E.
	\label{eq:solve_E}
\end{equation}
\textcolor{black}{Here, we approximate the $l_0$-minimization problem with an $l_1$-minimization problem to obtain a convex decoder \cite{tao11}-\!\cite{Tropp11}.}

Using the first block row of (\ref{eq:QR}), we can now compute $x[0]$ as follows:
\begin{equation}
	x[0] = R_1^{-1} Q_1^\top (Y- \hat{E}_{q,K}).
	\label{eq:QR1}
\end{equation}

The following result provides the conditions under which exists a unique solution to (\ref{eq:QR1}). The proof follows by using Lemma \ref{lem:CS} and the fact that the null space of $Q_2^\top$ is equal to the column space of $\Phi$.
\begin{lem} \label{lem:EC}
	There exists a unique solution $x[0] $ if all subsets of $2s$ columns of $Q_2 ^\top$ are linearly independent and $\Phi$ is full column rank. 
\end{lem}

Next, we propose a dynamic state estimator for secure state estimation of dynamic states $\omega_i$'s and $\theta_i$'s in microgrids.

\section{Secure State Estimation for Microgrids}\label{sec_nonlinear_estimation}
Let us assume that the time is slotted in time slot of size $\delta$. At each time slot $k$, the central controller receives the measured values of voltage phase angles, generators' internal angles, and rotors' speed. Let $y_i[k]$ denote the measurement vector received from bus $i\in\mathcal{N}$. Since the measurement units and communication paths from the measurement units to the central controller are not secured, $y_i[k]$ is subject to cyber attacks and corruptions:
\begin{align}
\label{measurement1}y_i[k] &= \begin{bmatrix} \theta_i[k]\\ \omega_i[k] \end{bmatrix}  + e_i[k],~\forall i\in\mathcal{N}_f^{(S)},\\
\label{measurement2}y_i[k] &= \theta_i[k] + e_i[k],~\forall i\in\mathcal{N}\setminus\mathcal{N}_f^{(S)},
\end{align}
where $e_i[k]$ represents either a corruption or an attack signal injected by a malicious agent. Notice that $e_i[k]\in\mathbb{R}^2$ for all $i\in\mathcal{N}_f^{(S)}$, and $e_i[k]\in\mathbb{R}$ for all $i\in\mathcal{N}\setminus\mathcal{N}_f^{(S)}$.

Our goal is to reconstruct $\theta_i[k]$ for all $i\in\mathcal{N}$ and $\omega_i[k]$ for all $i\in\mathcal{N}_f^{(S)}$ by using the received measurements. The only assumption concerning the corrupted sensors is the number of sensors that are attacked or corrupted due to failures. We do not assume the errors $e_i[k]$'s follow any particular model, i.e., the elements of $e_i[k]$ can take any value in $\mathbb{R}$. If the sensor at bus $i$ is not attacked and the communication flow from bus $i$ to the central controller is not corrupted, then necessarily the elements of $e_i[k]$ are zero.  

\subsection{Formulation of Secure State Estimation}
The microgrid dynamics and power flows can be described by the dynamical equations in (\ref{swing1k})-(\ref{load_bus}). We apply the forward Euler discretization scheme to the continuous-time dynamics in (\ref{swing1k})-(\ref{load_bus}) to obtain a discrete-time approximation, assuming constant discretization step $\delta$. We begin with the synchronous generators, and use the same approach for inverter-interfaced power supplies and loads.

\textbf{Synchronous Generators:} For each synchronous generator at fictitious bus $i\in\mathcal{N}_f^{(S)}$, we have
\begin{equation}
\begin{aligned}
&\theta_i [k+1] = \theta_i[k] + \delta (\omega_i[k] - \omega_0), \\
&\omega_i[k+1]	= \alpha_i \omega_i[k]+\eta_i P_i^m[k]+\beta_i\\
&\qquad\qquad\qquad\quad\quad+\sum_{j\in \mathcal{N}_i} f_{ij}(\theta_i[k],\theta_j[k]), \\
&P_i^m[k+1]=\kappa_i P_i^m[k]+\zeta_i[k] -\nu_i(\omega_{i}[k]+\begin{bmatrix} 0,1 \end{bmatrix}e_i[k]),\nonumber
\end{aligned}
\end{equation}
where $f_{ij}(\theta_i[k],\theta_j[k]) = \gamma_{ij}[k] \sin (\theta_i [k] - \theta_j [k]+\phi_{ij})$,
\begin{equation}
\begin{split}
\eta_i&=\delta/M_i,\\
\nu_i&=R_i \delta/\tau_i,\\
\kappa_i&=(1-\delta/\tau_i), \\
\alpha_i &= {(M_i-\delta D_i)}/{M_i},
\end{split}
\begin{split}
 \beta_i &= \delta D_i \omega_0/{M_i},\\
 \zeta_i[k]&=\delta(P_i^s[k]+R_i \omega_0)/\tau_i,\\
\gamma_{ij}[k] &= -{\delta V_i[k] V_j[k] |y_{ij}|}/{M_i}.
\end{split}		
\end{equation}
Notice that $\omega_{i,{\text{meas}}}[k]$ equals $\begin{bmatrix} 0,1 \end{bmatrix}y_i[k]$ which is equal to $(\omega_{i}[k]+\begin{bmatrix} 0,1 \end{bmatrix}e_i[k])$. Using (\ref{measurement1})-(\ref{measurement2}), we have
\begin{align}
	f_{ij}(\theta_i[k],\theta_j[k]) &= \gamma_{ij}[k] \sin (\theta_i [k] - \theta_j [k]+\phi_{ij})\nonumber \\
		 = &\gamma_{ij}[k] \sin(y_i[k] - e_i[k] - y_j[k] + e_j[k]+\phi_{ij}) \nonumber\\
		= &\gamma_{ij}^s [k] - \gamma_{ij}^s [k] e_{ij}^c  [k]  -  \gamma_{ij}^c  [k]  e_{ij}^s [k],
\end{align}
where
\begin{equation}
\begin{aligned}
\gamma_{ij}^s [k] &= \gamma_{ij}[k] \sin (y_i[k] - y_j[k]+\phi_{ij}),\\
\gamma_{ij}^c [k] &= \gamma_{ij}[k] \cos (y_i[k] -y_j[k]+\phi_{ij}),\\
e_{ij}^c[k] &\triangleq 1- \cos(e_i[k] - e_j[k]),\\
e_{ij}^s [k] &\triangleq \sin(e_i[k] - e_j[k]).
\end{aligned}
\end{equation}
Notice that the coefficients of $\gamma_{ij}^c[k]$ and $\gamma_{ij}^s[k]$ can be calculated using the measurement received from the local controllers.

Now, the state space model of the synchronous generator at fictitious bus $i\in\mathcal{N}_f^{(S)}$ can be represented by
\begin{equation}
\begin{aligned}
&x_i[k+1]= A_i x_i[k] +u_i[k] - \begin{bmatrix} 0 \\ \displaystyle H_i[k] \epsilon_i[k] \\ 0\end{bmatrix}-B_i e_i[k]\\
&y_i[k] = C_i x_i[k] + e_i[k]
\end{aligned}
\end{equation}
where the state vector $x_i[k]=\begin{bmatrix} \theta_i[k] , \omega_i[k], P_i^m[k] \end{bmatrix}^\top$, and
\begin{eqnarray} 
\begin{aligned}
&A_i=\begin{bmatrix} 1 & \delta & 0\\ 0 & \alpha_i & \eta_i\\ 0 & -\nu_i& \kappa_i \end{bmatrix}, B_i=\begin{bmatrix} 0& 0\\ 0 &0\\ 0& \nu_i\end{bmatrix}, C_i=\begin{bmatrix} 1 & 0 & 0\\ 0 & 1 & 0\end{bmatrix},\\
&u_i[k]=\begin{bmatrix} -\delta \omega_0, \beta_i+\sum_{j\in \mathcal{N}_i} \gamma_{ij}^s [k], \zeta_i[k] \end{bmatrix}^\top,\\
&\epsilon_i[k]=\begin{bmatrix} e_{i j_1}^c[k],\cdots,e_{i j_{n(i)}}^c[k],e_{i j_1}^s [k],\cdots,e_{i j_{n(i)}}^s [k]\end{bmatrix}^\top,\\
&H_i[k]=\begin{bmatrix} \gamma_{i j_1}^s[k],\cdots,\gamma_{i j_{n(i)}}^s[k],\gamma_{i j_1}^c[k],\cdots,\gamma_{i j_{n(i)}}^c[k] \end{bmatrix}.
\end{aligned}\nonumber
\end{eqnarray}
Notice that $n(i)$ denotes the number of neighbors of node $i$ in graph $\mathcal{G}(\mathcal{N},\mathcal{E})$, i.e., $n(i)=|\mathcal{N}_i|$, and that nodes $j_1,\cdots,j_{n(i)}$ represent the neighbors of node $i$ in $\mathcal{G}(\mathcal{N},\mathcal{E})$. More precisely, we have $H_i[k]\in\mathbb{R}^{1\times2n(i)}$ and $\epsilon_i[k]\in\mathbb{R}^{2n(i)}$. Next, by using the same approach, we obtain the state space model of inverter-interfaced power supplies and loads.


\textbf{Inverter-interfaced Power Supplies:} The state space model of the inverter-interfaced power supply at fictitious bus $i\in\mathcal{N}_f^{(I)}$ can be described by
\begin{equation}
\begin{aligned}
x_i[k+1] &=  x_i[k] +u_i[k] -  H_i[k] \epsilon_i[k],\\
y_i[k] &=  x_i[k] + e_i[k],
\end{aligned}
\end{equation}
where $x_i[k]=\theta_i[k]$ and $u_i[k]=\delta P_i^s[k]/{D_i}+\sum_{j\in \mathcal{N}_i} \gamma_{ij}^s [k]$. $H_i[k]$ and $\epsilon_i[k]$ are defined above.

\textbf{Loads:} The state space model of the load at bus $i\in\mathcal{N}^{(S)}\cup\mathcal{N}^{(I)}\cup\mathcal{N}^{(L)}$ can be described by
\begin{equation}
\begin{aligned}
x_i[k+1] &=  x_i[k] +u_i[k] -  H_i[k] \epsilon_i[k],\\
y_i[k] &= x_i[k] + e_i[k],
\end{aligned}
\end{equation}
where $x_i[k]=\theta_i[k]$ and $u_i[k]=-\delta P_i^d[k]/{D_i} +\sum_{j\in \mathcal{N}_i} \gamma_{ij}^s [k]$. $H_i[k]$ and $\epsilon_i[k]$ are defined above.

Consider an enlarged system composed of all the system dynamics:
\begin{equation}
\begin{aligned}
X[k+1] &=  A X[k] + U[k] - H[k] \epsilon [k] -B E[k],\\
Y[k] &= \begin{bmatrix} y_1[k]\\ y_2[k] \\ \vdots \\ y_{|\mathcal{N}|}[k] \end{bmatrix} = C X[k] + E[k],
\end{aligned}
\end{equation}
where
\begin{equation}
\begin{aligned}
&A \triangleq \text{blkdiag} \{ A_1,\cdots,A_m,I_{m+2n+l}\} \in \mathbb{R}^{(|\mathcal{N}|+2m)\times (|\mathcal{N}|+2m)},\\
&B \triangleq \text{blkdiag} \{ B_1,\cdots,B_m,0_{m+2n+l}\} \in \mathbb{R}^{(|\mathcal{N}|+2m)\times (|\mathcal{N}|+m)},\\
&C \triangleq \text{blkdiag} \{ C_1, \cdots, C_m, I_{m+2n+l} \}  \in \mathbb{R}^{(|\mathcal{N}|+m) \times (|\mathcal{N}|+2m)},\\
&X[k] \triangleq  \begin{bmatrix} x_1[k]^\top,\cdots,x_{|\mathcal{N}|}[k]^\top \end{bmatrix}^\top \in \mathbb{R}^{(|\mathcal{N}|+2m)},\\
&U[k] \triangleq \begin{bmatrix} u_1[k]^\top,\cdots,u_{|\mathcal{N}|}[k]^\top\end{bmatrix}^\top  \in \mathbb{R}^{(|\mathcal{N}|+2m)},\\
& \epsilon[k] \triangleq \begin{bmatrix} \epsilon_1[k]^\top,\cdots,\epsilon_{|\mathcal{N}|}[k]^\top\end{bmatrix}^\top \in \mathbb{R}^{2 \sum_{i\in\mathcal{N}} n(i) }, \\
&E[k] \triangleq  \begin{bmatrix} e_1[k]^\top,\cdots,e_{|\mathcal{N}|}[k]^\top \end{bmatrix}^\top \in \mathbb{R}^{|\mathcal{N}|+m},\\
& H[k] \triangleq  \text{blkdiag} \left\{ \begin{bmatrix} 0_{1\times 2n(1)} \\ H_1[k] \\0_{1\times 2n(1)} \end{bmatrix},\cdots,\begin{bmatrix} 0_{1\times 2n(m)}\\H_{m}[k] \\0_{1\times 2n(m)}\end{bmatrix},\right.\\
&\quad\left.,H_{m+1}[k],\cdots,H_{|\mathcal{N}|}[k] \right\}\in \mathbb{R}^{(|\mathcal{N}|+2m)\times 2 \sum_{i\in\mathcal{N}}  n(i) }.
 \nonumber
\end{aligned}
\end{equation}

Consider $K$ time slots $k= 0 , \cdots , K-1$, and collect the measurements corresponding the $K$ time slots in vector $\overline{Y}$ as follows:
\begin{equation}
\begin{aligned}
\overline{Y} &= \begin{bmatrix} Y[0]  \\ Y[1] - C   U[0]  \\ Y[2] - CA  U[0] - C  U[1]  \\ \vdots \\ Y[K-1] - C \sum_{k=0}^{K-2}  A^{K-2-k} U[k] \end{bmatrix}.
\end{aligned}
\end{equation}
$\overline{Y}\in \mathbb{R}^{K(|\mathcal{N}|+m)}$ can be rewritten in the following form:
\begin{equation}
\begin{aligned}
\overline{Y} &= \Phi X [0] + \Psi  \overline{E},
\end{aligned}
\end{equation}
where $\Psi = \begin{bmatrix} \Psi_1 & \Psi_2 \end{bmatrix}$ with $\Psi_1 \in \mathbb{R}^{K (|\mathcal{N}|+m)\times K (|\mathcal{N}|+m)}$ and $\Psi_2 \in \mathbb{R}^{K (|\mathcal{N}|+m)\times 2 K \sum_{i\in\mathcal{N}} n(i) }$, and
 \begin{align}
 \Psi_1 &= \begin{bmatrix}   I_{|\mathcal{N}|+m} &  0  &  \cdots  &\cdots  	\\
   			        -C B & I_{|\mathcal{N}|+m}&  \cdots  & \cdots	\\
			       -C  A B &  -C B & I_{|\mathcal{N}|+m}&\cdots &	\\
			         \vdots & \vdots  &  \ddots&\vdots\\
			         -C  A^{K-2} B &  \cdots &  -CB & I_{|\mathcal{N}|+m}\\
			    \end{bmatrix},\nonumber\\
\Psi_2 &= \begin{bmatrix}   0 &  0  &  \cdots   &\cdots 	\\
   			        -C H[0] & 0 &  \cdots  & \cdots	\\
			       -C  A H[0] &  -C H[1] & 0&\cdots &	\\
			         \vdots & \vdots  &  \ddots&\vdots\\
			         -C  A^{K-2} H[0] &  \cdots &  -CH[K-2] & 0\\
			    \end{bmatrix},\nonumber\\
\Phi  &= \begin{bmatrix} C ; C A ; C A ^2 ; \cdots ; C A^{K-1}\end{bmatrix}, \nonumber\\
\overline{E} &=    \begin{bmatrix} E[0]; \cdots; E[K-1] ; \epsilon[0] ;  \cdots ;\epsilon[K-1]\end{bmatrix}. \nonumber
\nonumber
\end{align}

The number of columns of the matrix $H[k]$ and the number of rows of the column vector $\epsilon[k]$ can be reduced from $2 \sum_{i\in\mathcal{N}}  n(i)$ to $2|\mathcal{E}|$ by using the following properties:
\begin{equation} \label{Eq:ec_es}
\begin{aligned}
	e_{ij}^c [k] &= 1- \cos( e_i[k] - e_j[k])  = e_{ji}^c [k],\\
	e_{ij}^s [k] &= \sin(e_i[k] - e_j[k]) =  - e_{ji}^s [k].
\end{aligned}
\end{equation}
Notice that we have $\gamma_{ij}[k]=\gamma_{ji}[k]$ and $\phi_{ij}=\phi_{ji}$. Hence, the dimension of $\Psi_2$ and $\overline{E}$ can also be reduced as follows:
\begin{equation}
\begin{aligned}
	\dim (\Psi_2) & \rightarrow {K (|\mathcal{N}|+m)\times 2 K |\mathcal{E}|},\\
	\dim (\overline{E}) & \rightarrow {K\left(|\mathcal{N}|+m+2 |\mathcal{E}|\right) }.
\end{aligned}
\end{equation}
We now choose $\Omega \in \mathbb{R}^{ (K|\mathcal{N}| - 2|\mathcal{N}| )\times K (|\mathcal{N}|+m)}$ which annihilates $\Phi$, i.e., $\Omega \Phi=0$. We then have:
\begin{equation}
	\widetilde{Y} = \Omega \overline{Y} = \Omega \widetilde{\Psi} \widetilde{E},
\end{equation}
where $\widetilde{\Psi}$ and $\widetilde{E}$ are the reduced $\Psi$ and $\overline{E}$.

In error correction, accurate decoding can be guaranteed if the coding matrix (i.e., $\Omega \tilde \Phi$) satisfies the Restricted Isometric Properties (RIP), which is often achieved  by randomly choosing a coding matrix {\it a priori}. In \cite{ctrl_sec9}, Theorem 1 provides a sufficient condition for perfect recovery of the system states against sensor attack and describes estimator design by using a state feedback controller. However, in the current setting, since there is a limitation to manipulate the coding matrix, we combine our secure estimator with a Kalman Filter (KF) to improve its practical performance. Since KF filters out both occasional estimation errors by the secure estimator and noisy measurement, a secure estimation based Kalman Filter improves practical performance \cite{ctrl_sec9}. Next, we numerically demonstrate the effectiveness of
the proposed estimators.


\section{Numerical Results}\label{num_sec}
\begin{figure}
\begin{center}
\includegraphics[width=2in]{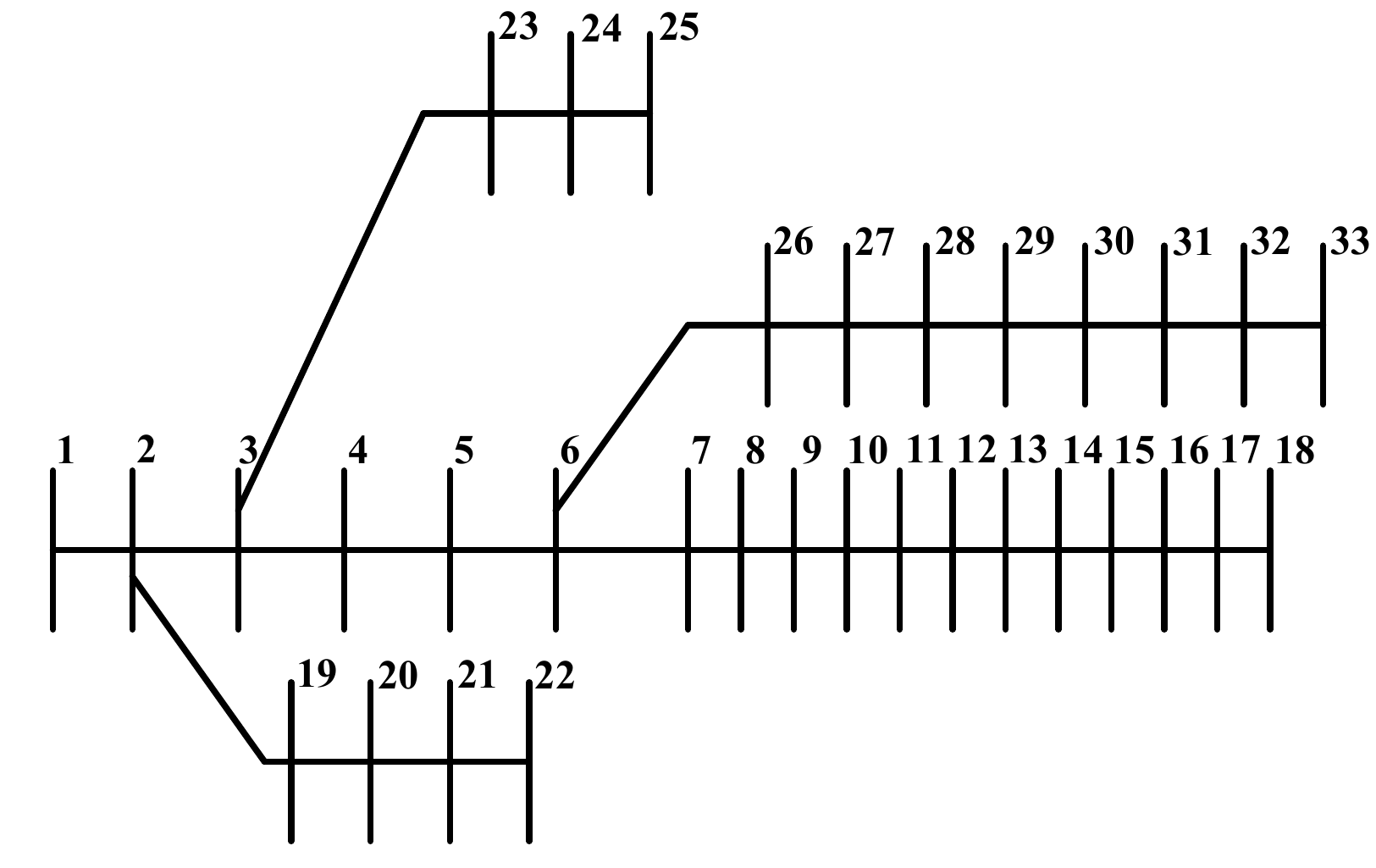}\caption{An AC microgrid comprising $m=3$ synchronous generators, $n=25$ inverter-interfaced power supplies, and $l=5$ load buses.}\label{fig:network}
\end{center}
\end{figure}

We consider an AC microgrid consisting of $m=3$ synchronous generators, $n=25$ inverter-interfaced power supplies, and $l=5$ load buses. The microgrid topology is shown in Fig. \ref{fig:network}, a modified IEEE 33-bus distribution system from \cite{testSystem}.  \textcolor{black}{Notice that in the augmented network,
the number of buses is 61. The synchronous generator buses are buses $\mathcal{N}^{(S)}=\{3,6,9\}$, the load buses without generation are buses $\mathcal{N}^{(L)}=\{1,2,14,22,25\}$, and the inverter-interfaced power supply buses are $\mathcal{N}^{(I)}=\{1,\cdots,33\}\setminus(\mathcal{N}^{(L)}\cup\mathcal{S}^{(L)})$. Further, we consider the fictitious buses for the synchronous generators to be buses 34,35, and 36, and the fictitious buses for the inverter-interfaced power supplies to be buses 37 to 61.}
The buses are connected via a set of electrical transmission lines. The microgird is connected to the grid via the tie-line connected to bus one. We take the turbine time constants to be $\tau_i=5$ s, the generator damping coefficients to be $D_i=2$, the inertia constants to be $M_i=10$, and the droop coefficients to be $R_i=9.5$ for all $i\in\mathcal{N}^{(S)}$. We further select the inverter-interfaced power supply droop coefficients to be $D_i=0.7$ for all $i\in\mathcal{N}^{(I)}$, and the constant frequency coefficients of the loads to be $D_i=0.1$ for all $i\in\mathcal{N}^{(L)}$.

We simulate the system for $t=20$ seconds with a discretization step of $\delta=1/60$ seconds. We select the nominal loads $P_i^d$'s randomly from the interval $[0,0.5]$~pu, and select the generation set-points $P_i^s$'s such that the system is balanced. Notice that computing the active power set-points $P_i^s$'s is not the subject of this study. Therefore, we only need to select the values of $P_i^s$'s such that the demand and supply are balanced. We further ignore reactive powers for synchronous generators, inverter-interfaced power supplies, and loads as they are unnecessary for the analysis presented in this study. Without loss of generality, we assume that voltage magnitudes $V_i[k]$'s are constant and equal to $1$~pu for all $k$ and $i=1,\cdots,33$.

We demonstrate the effectiveness of our secure state estimation method through simulations of the two types of attacks:
\begin{enumerate}
\item \textbf{Type A:} attacks targeted at synchronous generators, i.e., measurements $y_i[k]$'s, where $i \in \mathcal{N}^{(S)} \cup \mathcal{N}_f^{(S)}$, are corrupted.
\item \textbf{Type B:} attacks targeted at inverter-interfaced power supplies, i.e., measurements $y_i[k]$'s, where $i \in \mathcal{N}^{(I)} \cup \mathcal{N}_f^{(I)}$, are corrupted.
\end{enumerate}
For each of these attack types, we simulate three scenarios:
\begin{enumerate}
\item \textbf{Scenario 1:} There is no cyber attack on the microgrid.
\item \textbf{Scenario 2:} The microgrid is under cyber attack, and it is not protected by any secure state estimation.
\item \textbf{Scenario 3:} The microgrid is under cyber attack, and it is protected by the proposed secure state estimation.
\end{enumerate}

\subsection{Attack Type A: Synchronous Generator Attacks}\label{sec:attackA}

\begin{figure}
\centering
\includegraphics[scale=0.4]{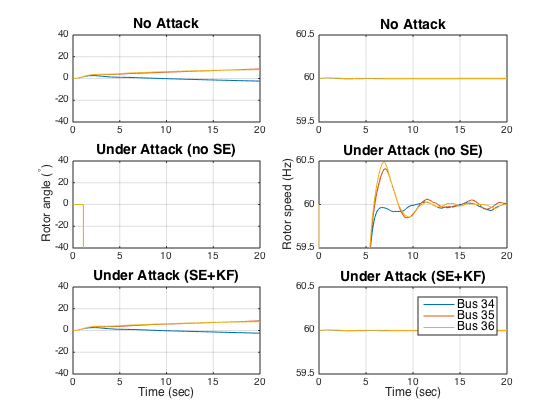}
\caption{Phase angles and rotor speeds of the synchronous generators under attack Type A in the three scenarios. }
\label{fig:case4_SE}
\end{figure}

The microgrid operator measures rotor angle from the three synchronous generator buses, and rotor angle and speed from the three fictitious synchronous generator buses. Hence, the operator has access to nine measurements that are subject to cyber attacks. We assume that from $t = 1.1$~seconds onwards, the attacker randomly chooses a set of five measurements out of the nine measurements and corrupts them with random signals at each time step. Fig.~\ref{fig:case4_SE} shows the simulation results for the three scenarios:  1) there is no attack (No Attack), 2) the microgrid system is under attack and there is no secure estimation (SE), and 3) the microgrid is under attack and the microgrid operator uses SE. Notice that in Fig.~\ref{fig:case4_SE}, we only show phase angles and rotor speeds of the three fictitious synchronous generator buses.

\begin{figure}
\centering
\includegraphics[scale=0.35]{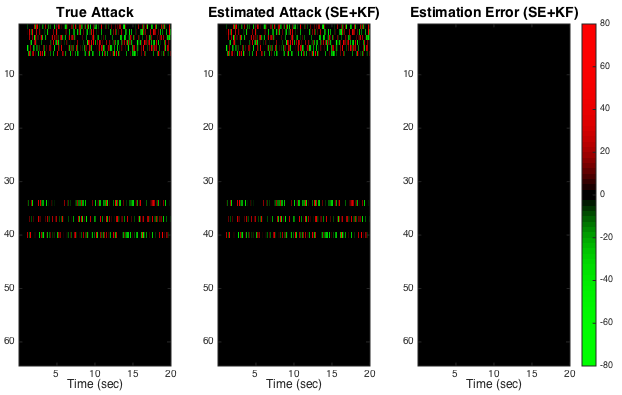}
\caption{True and estimated attack signals: The rows and columns correspond to measurements and time steps, respectively. In the subfigures, the color indicates the attack signal: red is a positive attack, green is a negative attack, and black is no attack. \textcolor{black}{Only measurements of the synchronous generators buses and their corresponding fictitious buses are corrupted by the attack. Notice that measurements 1 to 6 correspond to rotor angle and speed measurements from the fictitious synchronous generator buses, and measurements 34, 37 and 40 correspond to phase angle measurements from the synchronous generator buses.} In subfigure ``Estimation Error'', the black color indicates there is zero estimation error for all measurements at all times.}

\label{fig:case4_attack}
\end{figure}

In the microgrid without cyber attacks, the rotor speeds converge to 60~Hz after an initial transient period. As mentioned earlier, in Scenario 2, no secure estimation-based protection is implemented while the micorgrid operator uses the proposed secure state estimation to protect the system against cyber attacks in Scenario 3. Therefore, when the system is under cyber attack, both rotor angles and speeds cannot be estimated correctly in Scenario 2, as shown in Fig.~\ref{fig:case4_SE}. However, the operator can perfectly estimate both rotor angles and speeds in Scenario 3. Fig.~\ref{fig:case4_attack} shows the attack signal, secure estimator's estimated attack signal, and the estimation error. The results show that the secure estimator correctly estimates the attack signal throughout the simulation. Hence, the system's dynamics are identical to the system without any attacks.

\subsection{Attack Type B: Inverter-interfaced Power Supply Attacks}

\begin{figure}
\centering
\includegraphics[scale=0.4]{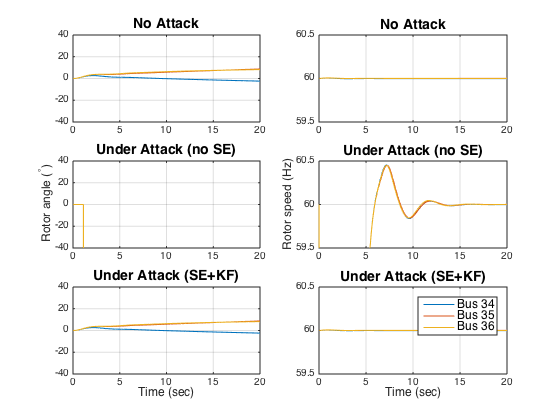}
\caption{Phase angles and rotor speeds of the synchronous generators under attack Type B in the three scenarios.}
\label{fig:case5_SE}
\end{figure}

\begin{figure}
\centering
\includegraphics[scale=0.35]{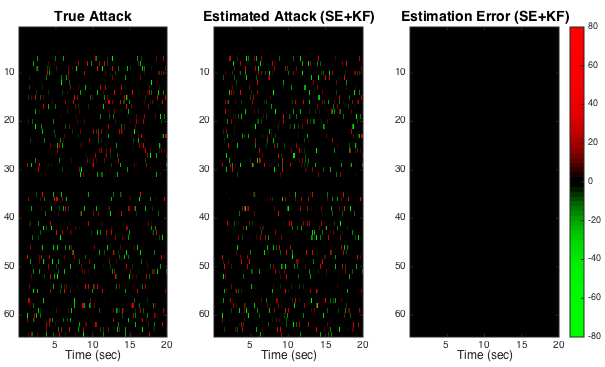}
\caption{True and estimated attack signals: The rows and columns correspond to measurements and time steps, respectively. In the subfigures, the color indicates the attack signal: red is a positive attack, green is a negative attack, and black is no attack. Only measurements of the inverter-interfaced power supply buses and their corresponding fictitious buses are corrupted by the attack. In subfigure ``Estimation Error'', the black color indicates there is zero estimation error for all measurements at all times.}
\label{fig:case5_attack}
\end{figure}

The microgrid operator measures phase angles from the inverter-interfaced power supply buses. Hence, the operator has access to 50 measurements that are subject to cyber attack: 25 correspond to phase angle measurements of the inverter-interfaced power supply buses and 25 correspond to the phase angle measurements of the corresponding fictitious buses. We assume that from $t = 1.1$~seconds onwards, the attacker randomly chooses a set of five measurements out of the 50 measurements and corrupts them with random signals at each time step. The simulation results are shown in Fig. \ref{fig:case5_SE} and Fig.~\ref{fig:case5_attack}. The results show that both phase angles and rotor speeds are severely affected when the system is under attack and no secure estimation-based protection is implemented. However, the microgrid operator can perfectly recover the attack signals and restore the system's normal dynamics as if there was no attack when secure state estimation is used.

\section{Conclusion}
We propose a secure state estimator for dynamic state estimation in AC microgrids under cyber physical attacks. We show that the microgrid operator can perfectly estimate the dynamic states in its AC microgrid using our estimator.


\begin{thebibliography}{00}
\bibitem{ref1} R. Lasseter, ``MicroGrids," {\em in IEEE Power Engineering Society Winter Meeting}, vol. 1, pp. 305-€"-308, 2002.

\bibitem{ref2} N. Hatziagyriou, H. Asano, R. Iravani, and C. Marnay, ``Microgrids," {\em IEEE Power and Energy Magazine}, vol. 5, no. 4, pp. 78--€"94, 2007.



\bibitem{ref_new_1} M. Ili\'c and S. Liu, {\em Hierarchical Power Systems Control: Its Value in a Changing Industry}. Springer, 1996.

\bibitem{ref_new_2}  P. Kundur, \textit{et al.}, ``Definition and classification of power system stability IEEE/CIGRE joint task force on stability terms and definitions," {\em IEEE Trans. on Power Systems}, vol. 19, no. 3, pp. 1387--1401, Aug. 2004.

\bibitem{ref_new_3}  F. C. Schweppe and J. Wildes, ``Power System Static-State Estimation, Part I: Exact Model," {\em IEEE Trans. on Power Apparatus and Systems}, vol. PAS-89, no. 1, pp. 120--125, Jan. 1970.

\bibitem{ref_new_4}  A. Abur and A. G. Exp\'osito, {\em Power System State Estimation: Theory and Application}. Marcel Dekker, Inc., 2004.

\bibitem{ref_new_12} J. Zhao, \textit{et al.}, ``Power System Dynamic State Estimation: Motivations, Definitions, Methodologies and Future Work," {\em IEEE Trans. on Power Systems}, 2019.


\bibitem{ref_new_5} H. Modir and R. A. Schlueter, ``A Dynamic State Estimator for Dynamic Security Assessment," {\em IEEE Trans. on Power Apparatus and Systems}, vol. PAS-100, no. 11, pp. 4644--4652, Nov. 1981.

\bibitem{ref_new_6} 
A. P. S. Meliopoulos, \textit{et al.}, ``Dynamic State Estimation-Based Protection: Status and Promise," {\em IEEE Trans. on Power Delivery}, vol. 32, no. 1, pp. 320--330, Feb. 2017.



\bibitem{ref_new8}  F. Pasqualetti, A. Bicchi, and F. Bullo, ``Consensus computation in unreliable networks: A system theoretic approach," {\em IEEE Trans. Autom. Control}, vol. 57, no. 1, pp. 90--104, Jan. 2012.

\bibitem{ref_new9}  F. Pasqualetti, F. Dorfler, and F. Bullo, ``Attack detection and identification in cyber-physical systems," {\em IEEE Trans. Autom. Control}, vol. 58, no. 11, pp. 2715--2729, Nov. 2013.

\bibitem{ref_new10}  K. Manandhar, X. Cao, F. Hu, and Y. Liu, ``Detection of faults and attacks including false data injection attack in smart grid using kalman
filter," {\em IEEE Trans. Control Netw. Syst.}, vol. 1, no. 4, pp. 370--379, 2014.



\bibitem{ref_new12}  S. Bi, and Y. J. Zhang, ``Graphical methods for defense against false-data injection attacks on power system state estimation," {\em IEEE Trans. Smart Grid}, vol. 5, no. 3, pp. 1216--1227, May 2014.

\bibitem{ref_new13}  L. Liu, M. Esmalifalak, Q. Ding, V. A. Emesih, and Z. Han, ``Detecting false data injection attacks on power grid by sparse optimization," {\em IEEE Trans. Smart Grid}, vol. 5, no. 2, pp. 612--621, March 2014.

\bibitem{ref_new14}  Y. Mo, R. Chabukswar, and B. Sinopoli, ``Detecting integrity attacks on scada systems," {\em IEEE Trans. Control Syst. Technol.}, vol. 22, no. 4, pp. 1396--1407, July 2014.

\bibitem{ref_new15}  T. Alpcan, and T. Basar, ``A game theoretic approach to decision and analysis in network intrusion detection," {\em in Proc. 42nd IEEE Conference on Decision and Control}, vol. 3, Dec 2003, pp. 2595--2600.

\bibitem{ref_new16} P. Srikantha and D. Kundur, ``A DER attack-mitigation differential game
for smart grid security analysis," {\em IEEE Trans. Smart Grid}, vol. 7, no. 3, pp. 1476--1485, May 2016.

\bibitem{ref_new19}  M. Basseville, I. V. Nikiforov \textit{et al.} , \textit{Detection of abrupt changes: theory
and application}. Prentice Hall Englewood Cliffs, 1993, vol. 104.

\bibitem{ref_new20}  Q. Jiao, H. Modares, F. L. Lewis, S. Xu, and L. Xie, ``Distributed $\ell_2$-gain
output-feedback control of homogeneous and heterogeneous systems," {\em Automatica}, vol. 71, pp. 361--368, 2016.


\bibitem{ref_new_7} S. Abhinav, H. Modares, F. L. Lewis, F. Ferrese, and A. Davoudi, ``Synchrony in Networked Microgrids Under Attacks," {\em IEEE Trans. on Smart Grid}, vol. 9, no. 6, pp. 6731--6741, Nov. 2018.


\bibitem{ctrl_sec8} H. Fawzi, P. Tabuada, and S. Diggavi, ``Secure estimation and control for
cyber-physical systems under adversarial attacks," {\em IEEE Trans. on Automatic Control}, vol. 59, no. 6, pp. 1454--1467, June 2014.


\bibitem{new2} Y. Shoukry and P. Tabuada, ``Event-Triggered State Observers for Sparse Sensor Noise/Attacks," {\em IEEE Trans. on Automatic Control}, no. 99, pp. 1--13, 2015.


\bibitem{ctrl_sec9}  Y. H. Chang, Q. Hu,  and C. J. Tomlin, ``Secure estimation based Kalman Filter for cyber physical systems against sensor attacks," {\em Automatica}, 95, pp.399-412, Sep. 2018.

\bibitem{ctrl_nonlin} Q. Hu, D. Fooladivanda, Y. H. Chang, and C. J. Tomlin, ``Secure State Estimation and Control for Cyber Security of the Nonlinear Power Systems," {\em  IEEE Trans. on Control of Network Systems}, vol. 5, no. 3, pp. 1310--1321, Sept. 2018.

\bibitem{new5} Y. Shoukry, P. Nuzzo, A. Puggelli, A. L. Sangiovanni-Vincentelli, S. A.
Seshia, and P. Tabuada, ``Secure state estimation for cyber physical
systems under sensor attacks: a satisfiability modulo theory approach,"
arXiv pre-print, Dec. 2014.


\bibitem{new6} M. Pajic, J. Weimer, N. Bezzo, P. Tabuada, O. Sokolsky, I. Lee,
and G. Pappas, ``Robustness of attack-resilient state estimators," {\em in
ACM/IEEE International Conference on Cyber-Physical Systems (ICCPS)}, 2014.




\bibitem{new10} S. Farahmand, G. B. Giannakis, and D. Angelosante, ``Doubly robust
smoothing of dynamical processes via outlier sparsity constraints," {\em IEEE
Trans. on Signal Processing}, vol. 59, no. 10, pp. 4529-–4543, Oct. 2011.

\bibitem{load_dyn1} X. Zhang, J. Chen, and C. Wang, ``Stability analysis of islanded microgrids with dynamic loads," {\em in Proc. IEEE 14th International Conference on Control, Automation, Robotics, and Vision}, pp. 1-6, 2016.

\bibitem{load_dyn2} A. Haddadi, A. Yazdani, G. Joos, and B. Boulet, ``A Generic load model for simulation studies of microgrids," {\em in Proc. IEEE Power \& Energy Society General Meeting}, pp. 1-5, 2013.


\bibitem{ctrl_dyn2} M. Yazdanian, and A. Mehrizi-Sani, ``Distributed Control Techniques in Microgrids," {\em in IEEE Trans. on Smart Grid}, vol. 5, no. 6, pp. 2901--2909, Nov. 2014.

\bibitem{ctrl_dyn3} A. Hooshyar, and R. Iravani, ``Microgrid Protection," {\em in Proc. of the IEEE}, vol. 105, no. 7, pp. 1332--1353, July 2017.

\bibitem{kundur_time} S. Liu, B. Chen, T. Zourntos, D. Kundur, and K. Butler-Purry, ``A Coordinated Multi-Switch Attack for Cascading Failures in Smart Grid," {\em IEEE Trans. on Smart Grid}, vol. 5, no. 3, pp. 1183--1195, May 2014.

\bibitem{structure_prev} A. R. Bergen, and D. J. Hill, ``A structure preserving model for power system stability analysis," {\em Power Apparatus and Systems, IEEE Transactions
on}, no. 1, pp. 25--35, 1981.

\bibitem{sauer} P. W. Sauer and M. A. Pai, \emph{Power System Dynamics and Stability}.
Upper Saddle River, NJ, USA: Prentice-Hall, 1998.

\bibitem{nerc} Available at \url{http://www.nerc.com/pa/Stand/Pages/CIPStandards.aspx}

\bibitem{Candes_Tao} E. Candes, J. Romberg, and T. Tao, ``Stable Signal Recovery from Incomplete and Inaccurate Measurements," Communications on pure and applied mathematics, vol. 59, no. 8, pp. 1207--1223, 2006.






\bibitem{David_Chang2} D. Hayden, Y. H. Chang, J. Goncalves, and C. Tomlin, ``Sparse network identifiability via Compressed Sensing," {\em Automatica}, vol. 68, pp. 9--17, Jun. 2016.

\bibitem{tao11} E. J. Candes, and T. Tao, ``Decoding by linear programming," {\em IEEE Trans. on Info. Theory}, vol. 51, no. 12, pp. 4203--4215, Dec. 2005.

\bibitem{david11} D. L. Donoho, and M. Elad, ``Optimally sparse representation in general (nonorthogonal) dictionaries via $l_1$ minimization," {\em in Proc. National Academy of Sciences}, pp. 2197--2202, 2003.


\bibitem{Michael11}  M. Elad, and A. M. Bruckstein, ``A generalized uncertainty principle and sparse representation in pairs of bases," {\em IEEE Trans. on Information Theory}, vol. 48, no. 9, pp. 2558--2567, 2002.

\bibitem{Gribonval11} R. Gribonval, and M. Nielsen, ``Sparse representations in unions of bases," {\em IEEE Trans. on Information Theory}, vol. 49, no. 12, pp. 3320--3325, 2003.

\bibitem{Tropp11} J.A. Tropp, ``Greed is good: algorithmic results for sparse approximation," {\em IEEE Trans. on Information Theory}, vol. 50, no. 10, pp. 2231--2242, 2004.



\bibitem{rotor1} P. Tripathy, S. C. Srivastava, and S. N. Singh, ``A Divide-by-Difference-Filter Based Algorithm for Estimation of Generator Rotor Angle Utilizing Synchrophasor Measurements," {\em in IEEE Transactions on Instrumentation and Measurement}, vol. 59, no. 6, pp. 1562--1570, June 2010.

\bibitem{rotor2} G. K. Venayagamoorthy, and R. G. Harley, ``MLP/RBF neural-networkbased
online global model identification of synchronous generator," {\em IEEE
Trans. Ind. Electron.}, vol. 52, no. 6, pp. 1685--1695, Dec. 2005.

\bibitem{rotor3} A. D. Angel, P. Geurts, D. Ernst, M. G. Glavic, and L. Wehenkel,
``Estimation of rotor angles of synchronous machines using artificial
neural networks and local PMU-based quantities," {\em Int. J. Neurocomput.},
vol. 70, no. 16--18, pp. 2668--2678, Oct. 2007.

\bibitem{rotor4} V. Venkatasubramanian, and R. G. Kavasseri, ``Direct computation of
generator internal states from terminal measurements," {\em in Proc. 37th
Annu. Hawaii Int. Conf. Syst. Sci.}, 2004.

\bibitem{testSystem} M. E. Baran, and F. F. Wu, ``Network reconfiguration in distribution systems for loss reduction and load balancing," {\em in IEEE Transactions on Power Delivery}, vol. 4, no. 2, pp. 1401--1407, Apr 1989.



\end{thebibliography}
\end{document}